\documentclass[iop]{emulateapj}
\usepackage{epsfig}
\usepackage{apjfonts}
\usepackage{amsmath,color,bm}
\usepackage{booktabs}
\usepackage[breaklinks,colorlinks,urlcolor=blue,citecolor=blue,linkcolor=blue]{hyperref}
\shorttitle{Search for the GW stochastic background from primordial curvature perturbations}
\shortauthors{Kapadia et al.}

\usepackage{natbib}
\usepackage{nameref}
\usepackage{graphicx}
\usepackage{graphics}
\usepackage[space]{grffile}
\usepackage{latexsym}
\usepackage{amsfonts,amsmath,amssymb}
\usepackage{url}
\usepackage[utf8]{inputenc}
\usepackage{fancyref}
\usepackage{hyperref}
\usepackage{xcolor}
\usepackage[normalem]{ulem}

\begin{document}


\title{Search for the stochastic gravitational-wave background induced by primordial curvature perturbations in LIGO's second observing run}


\author{Shasvath J. Kapadia$^1$}
\author{Kanhaiya Lal Pandey$^1$}
\author{Teruaki Suyama$^2$}
\author{Shivaraj Kandhasamy$^3$}
\author{Parameswaran Ajith$^{1,4}$}
\affiliation{$^1$~International Centre for Theoretical Sciences, Tata Institute of Fundamental Research, Bangalore 560089, India}
\affiliation{$^2$Department of Physics, Tokyo Institute of Technology, 2-12-1 Ookayama, Meguro-ku, Tokyo 152-8551, Japan}
\affiliation{$^3$Inter-University Centre for Astronomy and Astrophysics, Pune 411007, India}
\affiliation{$^4$~Canadian Institute for Advanced Research, CIFAR Azrieli Global Scholar, MaRS Centre, West Tower, 661 University Ave, Toronto, ON M5G 1M1, Canada}



\begin{abstract}
Primordial density perturbations in the radiation-dominated era of the early Universe are expected to generate stochastic gravitational waves (GWs) due to nonlinear mode coupling.
In this \emph{Letter}, we report on a search for such a stochastic GW background in the data of the two LIGO detectors during their second observing run (O2). We focus on the primordial perturbations in the range of comoving wavenumbers $10^{16}-10^{18}~{\rm Mpc}^{-1}$ for which the stochastic background falls within the detectors' sensitivity band. We do not find any conclusive evidence of this stochastic signal in the data, and thus place the very first GW-based constraints on the amplitude of the power spectrum at these scales. We assume a lognormal shape for the power spectrum and Gaussian statistics for the primordial perturbations, and vary the width of the power spectrum to cover both narrow and broad spectra. Derived upper limits ($95\%$) on the amplitude of the power spectrum are $0.01-0.1$.  
As a byproduct, we are able to infer upper limits on the fraction of the Universe's mass in ultralight primordial black holes ($M_\mathrm{PBH} \simeq 10^{-20}-10^{-19}M_{\odot}$) at their formation time to be $\lesssim 10^{-25}$. 

\end{abstract}

\maketitle

\section{Introduction:}\label{para:introduction}

Cosmological observations have revealed that all the structures in the present 
Universe originate from the primordial density perturbations (equivalently, curvature perturbations).
According to the theory of inflation, which constitutes a pillar of modern cosmology, the primordial perturbations are created by the amplification of the quantum fluctuations of the scalar fields during inflation and existed over a wide range of length scales from meter scale up to at least the Hubble horizon scale \citep{Liddle-Lyth}.
Knowledge of the primordial perturbations is crucial to 
test inflation models and physics of the early Universe.

Observations of the cosmic microwave background (CMB) and 
the large-scale structure have successfully measured the power spectrum of the primordial perturbations on large scales as 
${\cal P}_\zeta \approx 2\times 10^{-9}$~\citep{Aghanim:2018eyx} with
a small scale-dependence.
However, much less is known of the primordial perturbations 
on smaller scales. 
At ${\cal O}(0.1 {\rm kpc})$, the non-detection
of the CMB spectral distortion places an upper limit of
${\cal P}_\zeta \lesssim 10^{-4}$ (see \cite{Chluba:2019kpb} and references therein).
Success of the big bang nucleosynthesis provides 
${\cal P}_\zeta \lesssim 10^{-2}$ for a range 
$0.01 {\rm kpc}\sim 0.1 {\rm kpc}$ \citep{Jeong:2014gna, Nakama:2014vla, Inomata:2016uip}. 
Non-detection of primordial black holes (PBHs) yields a similar
level of constraints ${\cal P}_\zeta \lesssim 10^{-2}$
for a wide range of scales (e.g. \cite{Allahverdi:2020bys}).

Stochastic gravitational waves (GWs), which is a target of this \emph{Letter}, 
have been attracting considerable interest recently as a powerful probe of the primordial perturbations (e.g. \cite{Inomata2019}).
At the second order in the cosmological perturbation, the mode-mode couplings of the primordial curvature perturbations induce a stochastic GW background \citep{Tomita:1967sv, Matarrese:1993zf}
\footnote{Although some earlier work had suggested the gauge dependence of the induced stochastic background of GWs, more recent work demonstrates that this is not the case. See \cite{InomataTerada2020} and references therein.}.
\cite{Ananda:2006af} and \cite{Bugaev2010, Bugaev2011} suggested that future GW detectors can be used to constrain the primordial perturbations
on very small scales.
\cite{Saito:2008jc} pointed out that GW observations can constrain the PBHs as dark matter candidates. 
\cite{Inomata2019} provides a summary of 
the expected constraints on the small-scale
primordial perturbations by the current/planned GW observations. 
Although there are many theoretical or observational-prospect studies on such stochastic GWs, 
no observational test using real GW data have been given in the literature.

In this \emph{Letter}, following our previous paper \citep{Kapadia2020} that explored the detection prospects for the isotropic stochastic GWs induced by the primordial perturbations, 
we report the results of the very first search for this signal in LIGO data from the second (O2) observing run \citep{GW_Open_Data}. This stochastic  background in LIGO's sensitive band corresponds to the primordial perturbations in the
comoving wavenumber $10^{16}~{\rm Mpc}^{-1} \lesssim k \lesssim 10^{18}~{\rm Mpc}^{-1}$. 
In the following analysis, we assume that the power spectrum of the primordial curvature perturbations has a lognormal shape 
defined by Eq.~(\ref{eq:powerspectrum})
which is characterized by three parameters: $A$ (amplitude), 
$k_0$ (comoving wavenumber at the peak of the power spectrum), and $\sigma$ (width).
We also assume that the primordial curvature perturbations obey Gaussian statistics.
Our analysis can be easily extended for other shapes of the power spectrum
and the non-Gaussian primordial perturbations.
We employ the cross-correlation search which is optimal for stationary and isotropic backgrounds that obey Gaussian statistics \citep{Romano2017, Christensen1992, Christensen2018, Allen1999}. 
Making use of the cross-correlation data released by the LIGO-Virgo collaboration 
from O2 \citep{O2_cross_correlators, O2_stochastic_search}, we estimate signal-to-noise ratios (SNRs) on a $k_0 - \sigma$ grid corresponding to the range of the comoving wavenumbers mentioned above, 
and a range of widths spanning both narrow and broad power spectra ($0.01 \leq \sigma \leq 10$).

We do not find any conclusive evidence for the presence of this GW background in the data (all SNRs $\lesssim 2.7$) 
\footnote{Note that our search targets a different stochastic GW spectrum (pertaining to a lognormal power spectrum and a different physical origin), which is not included in the search performed by the LIGO-Virgo collaboration~\citep{O2_stochastic_search}.}.
We therefore place upper limits on the amplitude of the curvature power spectrum using Bayesian parameter estimation where the likelihood is constructed from the cross-correlation 
and the assumed model of the stochastic background \citep{Mandic2012}. We find that $95 \%$ upper limits on the power spectrum amplitude 
span about $0.01 - 0.1$ for the majority of the parameter space considered. 

As a byproduct of the derived upper limits, we are able to constrain the PBH abundance in the mass range $10^{-20} - 10^{-19}M_{\odot}$ at the time of their formation.
The existing upper limits \citep[and the references therein]{Carr1_2017, Carr2_2017}, rely on the effects of Hawking radiation from evaporating black holes \citep{Hawking1975}. On the other hand, our constraints on the fraction $\beta^\prime$ of the Universe’s mass in the form of these ultralight PBHs at their formation time ($\beta^\prime \propto \rho_\mathrm{PBH}/\rho$; see, e.g., \cite{Carr2020})  are independent of the existence of Hawking radiation. For a narrow mass range, our constraints are comparable to or better than the existing constraints, if we assume narrow primordial power spectra ($0.01 \lesssim \sigma \lesssim 0.5$). These ultralight PBHs are expected to have evaporated by Hawking radiation by the current cosmic age. We show that, even if they have not, they would constitute only a very small fraction of the dark matter ($f_{\mathrm{PBH}} \equiv \Omega_\mathrm{PBH}/\Omega_\mathrm{DM}$ as low as $10^{-15} - 10^{-5}$). 

\section{Search for the stochastic GW background:}\label{para:method}

The isotropic stochastic GW background can be described in terms of the energy density fraction $\Omega_{\mathrm{GW}}$ per logarithmic frequency bin: 
%
\begin{equation}
\Omega_{\mathrm{GW}}(f) = \frac{1}{\rho_c}\frac{d\rho_{\mathrm{GW}}}{d\log(f)},
\end{equation}
where $\rho_{\mathrm{GW}}$ is the energy density of GWs and $\rho_c$ the critical energy density required for a flat Universe. If the GWs are sourced by scalar-tensor mode couplings in primordial curvature perturbations, $\Omega_{\mathrm{GW}}(f)$  would depend on the shape of the curvature power spectrum. Here, we assume the power spectrum to be of log-normal shape, parametrized by the amplitude $A$, central wave number $k_0$ and width $\sigma$~\footnote{The relation between the log-normal curvature power spectrum and the GW background is complicated and does not in general have a closed form. See, for example, \cite{Kapadia2020}, for a summary, and \cite{Wang2019, Kohri2018} for details.}:
\begin{equation}
P_{\zeta}(k) = \frac{A}{\sqrt{2\pi} \sigma}\exp\left(-\frac{\log^2(k/k_0)}{2\sigma^2} \right)
\label{eq:powerspectrum}
\end{equation}
where $k$ is the comoving wave number that sets the spatial scale. Since $k_0$ depends on the PBH mass-scale $M_{\mathrm{PBH}}$ \citep{Kohri2018, Inomata2019}, we can also use $M_{\mathrm{PBH}}$ to parametrize the power spectrum instead of $k_0$. 

The log-normal distribution is a natural choice (and often adopted in the literature) for the shape of the curvature power-spectrum, in the absence of the knowledge of its true shape. This shape conveniently encompasses an arbitrarily large range of central wavenumber scales, as well as widths which span both narrow and broad spectra.

The search for a stationary, Gaussian, unpolarized, and isotropic stochastic GW background involves the calculation of the following cross-correlation statistic $\hat{C}(f)$ from the data of two detectors~\citep{O2_stochastic_search} \footnote{Strictly speaking, Eq.~\eqref{eq:cc} should be interpreted as an average over multiple frequency bins $\Delta f$, where the cross-correlator in each bin is given by: $\hat{C}(f) = \frac{2}{T\Delta f}\int^{f+\Delta f/2}_{f-\Delta f/2} \frac{\mathrm{Re}\left[\tilde{s}_1^*(f')\tilde{s}_2(f')\right]}{\gamma_T(f')S_0(f')}df'$. For the O2 stochastic search, $\Delta f = 0.03125$ Hz, $T = 192$ sec., and the total livetime 
was $99$ days \citep{O2_stochastic_search}.}: 
\begin{equation}
\hat{C}(f) = \frac{2}{T} \frac{\mathrm{Re}[\tilde{s}_1^\star(f) \, \tilde{s}_2(f) ]}{\gamma_\mathrm{T}(f) \, S_0(f)},
\label{eq:cc}
\end{equation}
where $\tilde{s}_i(f)$ are the Fourier transforms of the time series data $s_i(t)$ from detector $i = \{1, 2\}$, $T$ is the duration of the data used to compute the Fourier transform, $\gamma_\mathrm{T}(f)$ is a geometric factor, called the overlap reduction function, that depends on the relative orientation of the detectors, while $S_0(f)$ is the spectral shape for a stochastic GW background with a flat $\Omega_{\mathrm{GW}}(f)$ (see, for e.g., \cite{O2_stochastic_search}). The expectation value and the variance of $\hat{C}(f)$ are given by:
\begin{equation}
\label{cross-correlator}
\langle \hat{C}(f) \rangle = \Omega_{\mathrm{GW}}(f), ~~~~ \sigma^2_C(f) \approx \frac{1}{2T\Delta f}\frac{P_1(f) P_2(f)}{\gamma_\mathrm{T}^2(f) S_0^2(f)}, 
\end{equation}
where $P_i(f)$ are the one-sided  power spectral density of the noise in the two detectors (assumed to be Gaussian) and $\Delta f$ is the frequency resolution of the discrete Fourier transform. 

Given the cross correlation $\hat{C}(f)$ and a signal model $\Omega_{\mathrm{GW}}(f)$, an optimal estimator $\hat{\Omega}_{\mathrm{ref}}$ for the signal, and its variance $\sigma^2_{\Omega}$, can be computed as the following weighted sums over the frequency bins $j$~\citep{O2_stochastic_search}:
\begin{eqnarray}
\hat{\Omega}_{\mathrm{ref}} = \frac{\sum_j w(f_j)^{-1}\hat{C}(f_j)\sigma_C^{-2}(f_j)}{\sum_j w(f_j)^{-2}\sigma_C^{-2}(f_j)}, ~~~
\sigma^{2}_{\Omega} = \frac{1}{\sum_j w(f_j)^{-2}\sigma_{C}^{-2}(f_j)},
\end{eqnarray}
where $w(f) := {\Omega_{\mathrm{GW}}(f_{\mathrm{ref}})}/{\Omega_{\mathrm{GW}}(f)}$ is a weight function and $f_{\mathrm{ref}}$ is a reference frequency which is set to $21$ Hz. 
The SNR of this estimator is $\hat{\Omega}_{\mathrm{ref}}/\sigma_{\Omega}$; it is therefore independent of the choice of $f_{\mathrm{ref}}$. When calculated from stationary Gaussian noise, SNR will be distributed according to a standard normal distribution. Thus, it can be directly interpreted as the significance of the signal detection in stationary Gaussian noise (in terms of Gaussian standard deviations). 


We can also compute the Bayesian posteriors of the signal parameters $\Theta := \lbrace A, \sigma, k_0 \rbrace$  from the cross correlation $\hat{C}(f)$. For stationary Gaussian noise, the  likelihood for $\hat{C}(f)$ is~\citep{Mandic2012}:
\begin{eqnarray}
p(\hat{C} \mid \Theta) &\propto& \prod_j \exp \left(\frac{-[\hat{C}(f_j) - \Omega_{\mathrm{GW}}(f_j; \Theta)]^2}{2\sigma_C^2(f_j)}\right). 
\label{likelihood}
\end{eqnarray}
We first fix $k_0$ and $\sigma$ and compute the posterior on $A$ assuming a suitably chosen prior. We then repeat this calculation over a grid of $k_0 - \sigma$. From the posterior distribution $p(A~|~\hat{C}, \sigma, k_0)$, we calculate $95\%$ upper limits on $A$, which can be used to derive an upper limit on $f_{\mathrm{PBH}}$ as done in \cite{Kapadia2020, Wang2018, Wang2019, Inomata2019}.

%
%

\section{Results:}\label{para:results}

\begin{figure}
\includegraphics[width=1.1\linewidth]{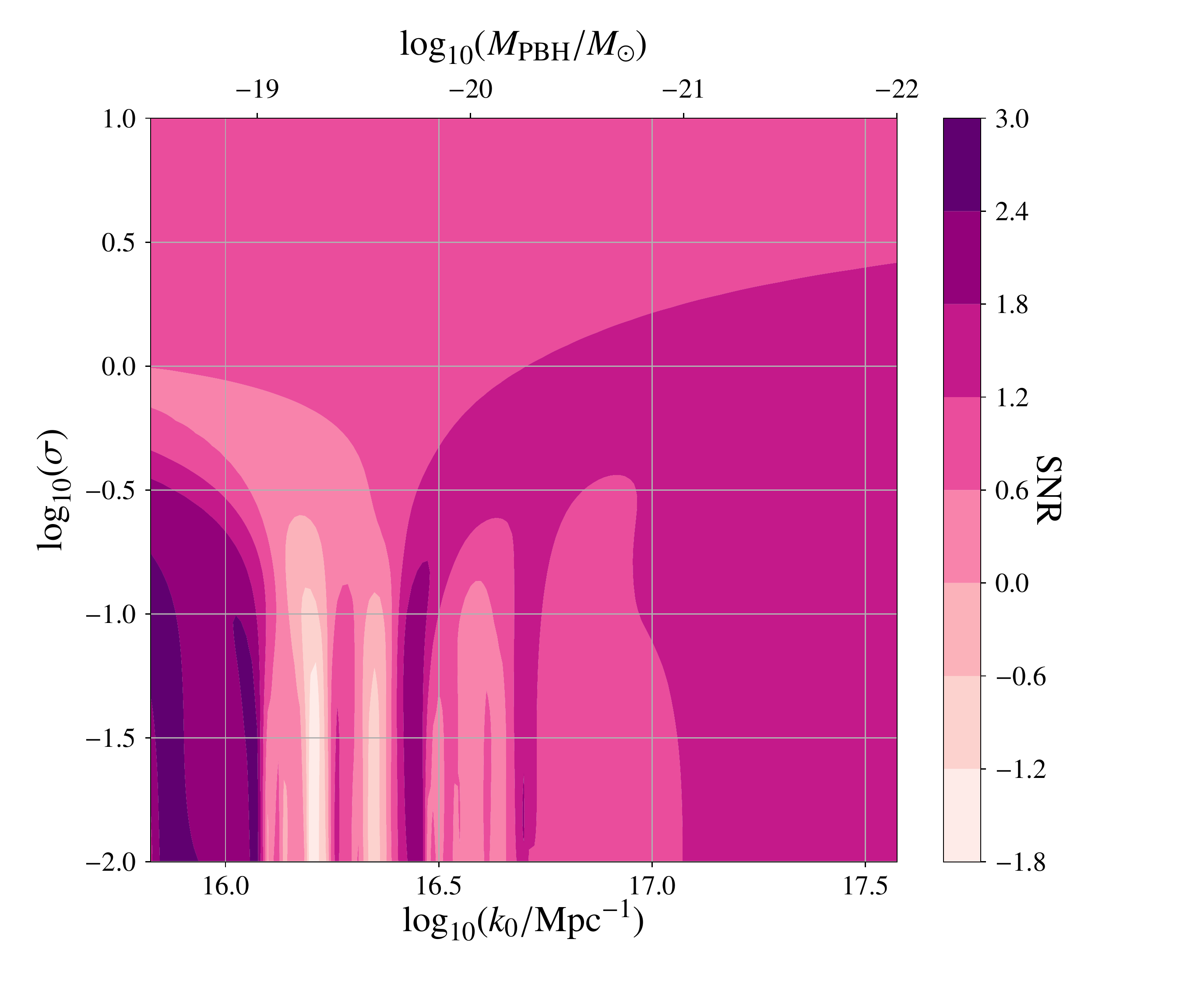}
\caption{Search SNR  evaluated on a grid of $\log_{10}(k_0/\mathrm{Mpc}^{-1})-\log_{10}(\sigma)$. The SNRs don't exceed $\sim 2.7$; we therefore do not find any conclusive evidence of a signal consistent with the $\Omega_\mathrm{GW}$s pertaining to the model parameter grid considered here.} 
\label{snr}
\end{figure}

\begin{figure*}[tbh]%
    \includegraphics[height=7.3cm]{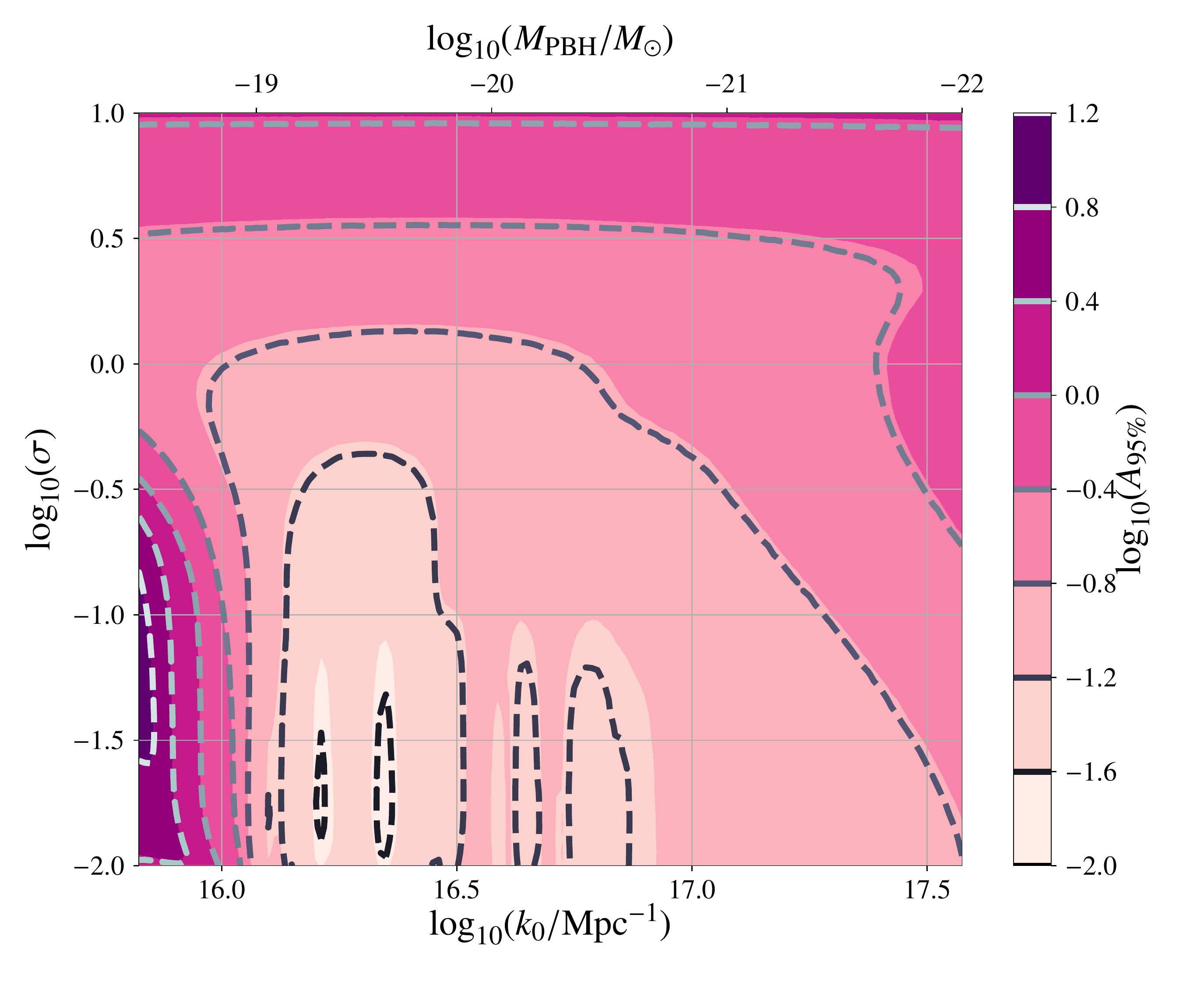}
    \includegraphics[height=7.3cm]{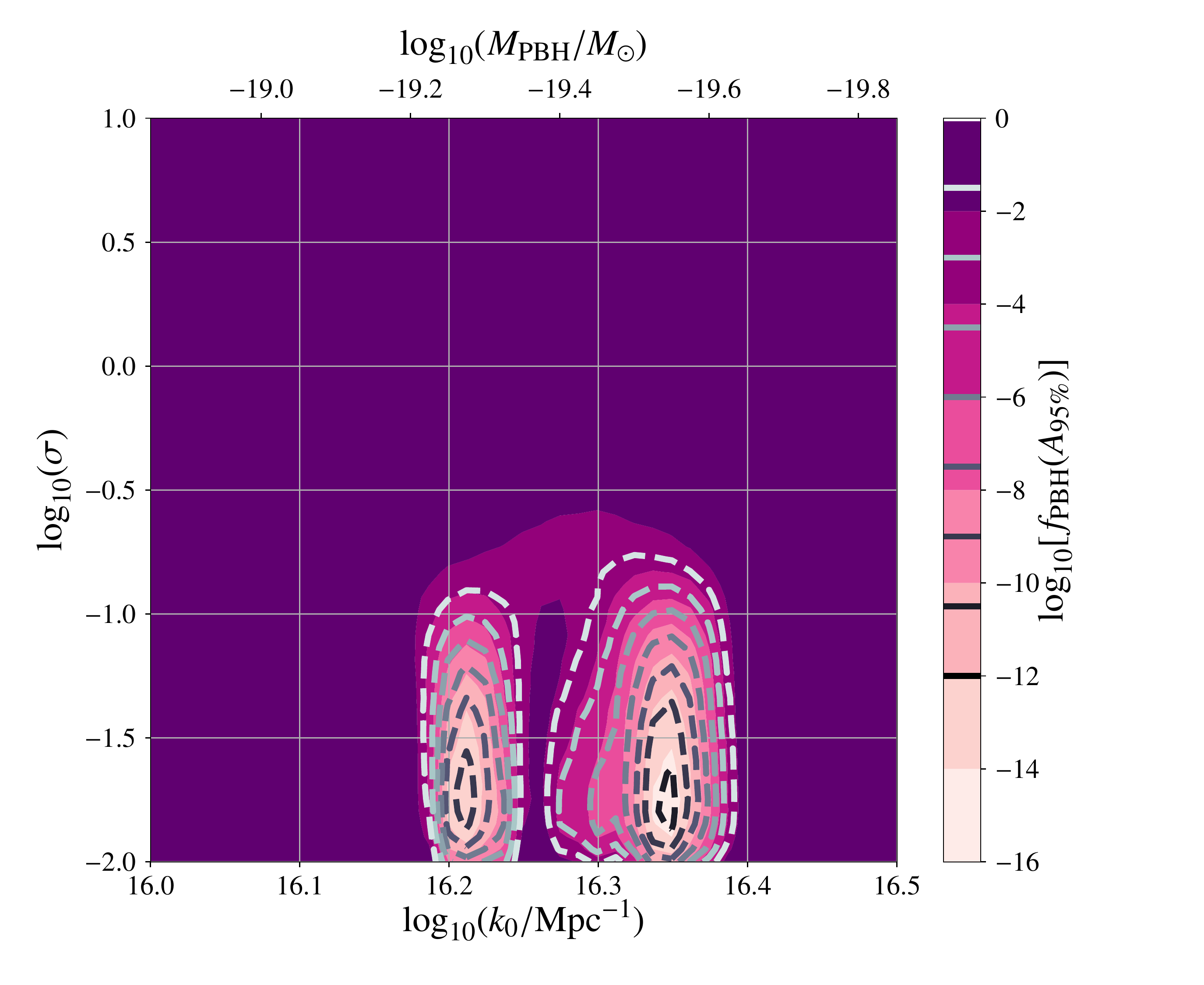} 
    \caption{{\it Left Panel}: Upper limits on the amplitude $A$ derived from the posterior distribution $p(A~\mid \hat{C}, \sigma, k_0)$, for a uniform (dashed contour lines) and log-uniform (filled contour areas) prior, on an $\log_{10}(k_0/\mathrm{Mpc}^{-1})-\log_{10}(\sigma)$ grid. The upper limits from the log-uniform prior are marginally tighter than those from the uniform prior, and can be as low as $\sim 0.01$. {\it Right Panel}: Upper limits on $f_{\mathrm{PBH}}$ derived from the upper limits on $A$. While non-trivial limits can only be placed for a narrow range of masses and a narrow range of $\sigma$'s, they can be as stringent as $\sim 10^{-15}$. These are stronger than those based on Hawking radiation, but only for certain small values of $\sigma$. }%
    \label{contour_ul}%
\end{figure*}

We evaluate the optimal estimator $\hat{\Omega}_{\mathrm{ref}}$, its variance $\sigma^2_{\Omega}$, and the SNR = $\hat{\Omega}_{\mathrm{ref}}/\sigma_{\Omega}$ on the model $\log_{10}(k_0/\mathrm{Mpc}^{-1})-\log_{10}(\sigma)$ parameter grid, with $k_0$ spanning $\sim 10^{16}-10^{18} \mathrm{Mpc}^{-1}$ ($M_{\mathrm{PBH}} \sim 10^{-22} - 10^{-18.5}M_{\odot}$), and $\sigma$ spanning $0.01-10$. We use the cross-correlation data $\hat{C}(f_j)$ and their variances $\sigma^2_C(f_j)$ released by the LIGO-Virgo collaboration for O2 \citep{O2_cross_correlators, O2_stochastic_search}. The results of the search are summarized in Fig.~\ref{snr}. We find no conclusive evidence of a signal: The maximum SNR over the search parameter space is $\sim 2.7$~\footnote{Note that this is not significant enough to even claim a tentative evidence of signal, as the trials factor for repeating the search over different signal parameter values is not included here. When the trials factors are included the significance of this detection is going to be much less than $2.7\sigma$. The lack of a tentative evidence is further confirmed when we evaluate the Bayes factor -- the ratio of the marginalized likelihoods under signal/noise hypothesis over the parameter space probed in Fig.~\ref{snr} -- whose value is found to be $\sim 1$.}.

We test our analysis on simulated values of $\hat{C}(f_j)$  for stationary Gaussian noise and find that searching for the stochastic background over the same grid of parameters yields SNRs that are consistent with that reported in Fig.\ref{snr}. We also evaluate the mean and standard deviation of the distributions of  SNRs estimated from several independent realizations of Gaussian noise. With increasing number of noise realizations, the mean (standard deviation) of the distributions approach zero (unity), as expected. This confirms that applying the cross-correlation method described in the previous section to searching for the stochastic backgrounds, pertaining to both narrow and broad power spectra, does not produce any unexpected biases.

We then estimate Bayesian posterior distributions on the amplitude, $p(A~\mid \hat{C}, \sigma, k_0)$ over the model parameter grid mentioned above, from the likelihood described in Eq.~\eqref{likelihood}, and two choices of prior, one uniform in $A$, and the other uniform in $\log_{10}(A)$. The upper limits derived from these posteriors are presented in Fig.~\ref{contour_ul} (left plot). We are able to constrain the amplitude to values as low as $\sim 0.01$ at $95\%$ confidence for certain $k_0 - \sigma$ values. Fig.~\ref{select_ul} (top panel) compares the upper limits on $A$ (for certain fiducial values of $\sigma  = 0.01, 5$) with existing ones from other experiments, including from GWs from compact binary coalescences. 

From the upper limits on the amplitude $A$, upper limits on $f_{\mathrm{PBH}}$ can be estimated, neglecting Hawking radiation (see, for e.g., \cite{Kapadia2020}). As shown in \cite{Kapadia2020, Kohri2018, Wang2019}, $f_{\mathrm{PBH}}$ is highly sensitive to changes in the amplitude; a change of a factor of $2$ could result in a change of many orders of magnitude in $f_{\mathrm{PBH}}$. The results of the conversion from upper limits on $A$ to upper limits on $f_{\mathrm{PBH}}$ on the model parameter grid $M_{\mathrm{PBH}} - \sigma$ are summarized in Fig.~\ref{contour_ul} (right plot). The $95\%$ upper limits on $f_{\mathrm{PBH}}$ are rather weak for a large portion of the parameter space considered. Nevertheless, for certain mass-scales between $10^{-20}-10^{-19} M_{\odot}$ and narrow spectra, upper limits can be as stringent as $10^{-15}-10^{-5}$. 

While the notion of these ultralight PBHs constituting even a fraction of the dark matter in the current cosmological epoch needs to neglect the effect of Hawking evaporation, our results can constrain the fraction $\beta'$ of the Universe's mass in the form of these PBHs at their formation time, independent of the (non)existence of Hawking radiation. Fig.~\ref{select_ul} (lower panel) compares our constraints on $\beta'$ with existing ones from other experiments. Note that our constraints fall in a mass-range where existing constraints assume Hawking radiation. For a narrow range of PBH masses, our constraints assuming $\sigma = 0.01, 0.1$ are comparable, and sometimes marginally stronger than the non-GW ones. These constraints can get significantly stronger for other $\sigma$ values (see Fig.~\ref{contour_ul}), but again limited to a narrow mass-range.

\begin{figure*}[tbh]%
    \vspace{-1cm}
    \includegraphics[width=17.6cm]{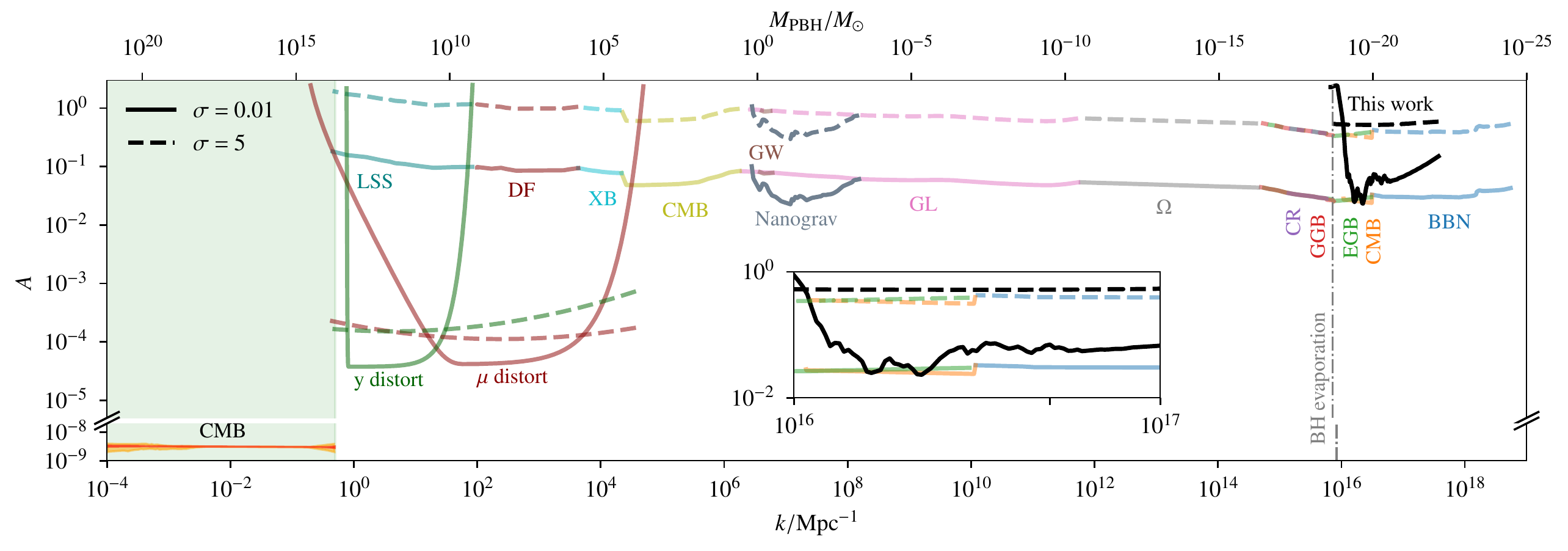}
    \includegraphics[width=17.7cm]{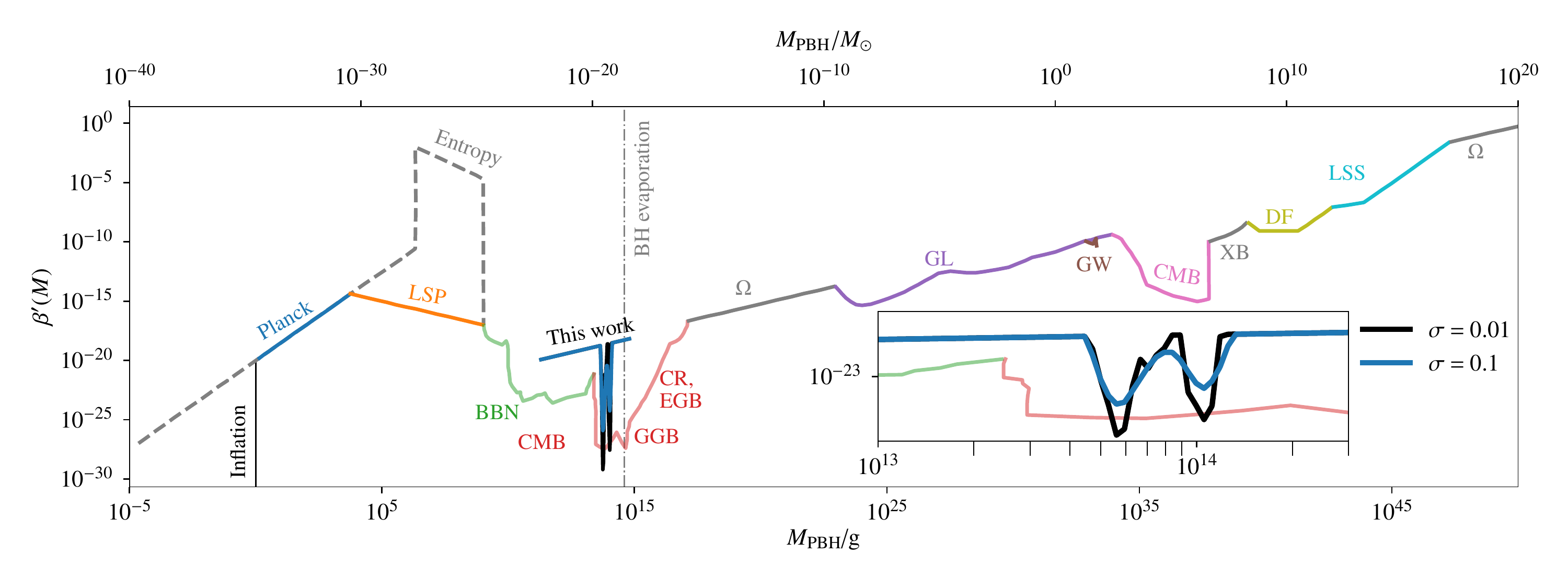}
        \caption{{\it Top Panel}: Upper limits (95\%) on the amplitude $A$ of the curvature power spectrum (assuming a log-uniform prior and two fiducial values 0.01 and 5 for $\sigma$), along with upper limits  from other experiments \citep{Carr2020}. The dashed vertical line corresponds to PBH masses below which Hawking radiation should have caused PBHs to evaporate by the current cosmic age. For $\sigma = 0.01$, the constraints from this work are marginally better than existing constraints, for a narrow mass range. For $\sigma = 5$, the constraints are marginally worse. Note however, that our constraints are Hawking-radiation independent, as opposed to other constraints in the same mass-range. {\it Bottom Panel}: Upper limits on the abundance of PBHs at the time of their formation, $\beta'$, for $\sigma = 0.01, 0.1$, along with constraints from other experiments \citep{Carr2020}. While non-trivial limits, corresponding to $f_{\mathrm{PBH}} < 1$, can only be placed for a narrow range of masses, they are comparable and can even be stronger than existing constraints. (Abbreviations: {\it LSP}: Lightest supersymmetric particle, {\it BBN}: Big bang nucleosynthesis, {\it CMB}: Cosmic microwave background, {\it GGB}: Galactic gamma-ray background, {\it EGB}: Extra-galactic photon background, {\it CR}: Cosmic rays, {\it $\Omega$}: $f_{\mathrm{PBH}} = 1$, {\it GL}: Gravitational lensing, {\it GW}: GW-based limits from LIGO-Virgo data, {\it XB}: X-ray background, {\it DF}: Dynamical friction, {\it LSS}: Large scale structure; see \cite{Carr2020} for details. {\it $\mu$ distort}: $\mu$ distortion, {\it $y$ distort}: $y$ distortion; (deviations of CMB energy spectrum from black-body spectrum); see \cite{Chluba2012} and \cite{Fixsen1996} for details. {\it NANOGrav}: GW-based limits from NANOGrav (pulsar-timing methods) 11-year data set; see \cite{Nanograv} for details.)}
   \label{select_ul}%
    \label{select_ul}%
\end{figure*}

\section{Summary and outlook:}\label{para:conclusion}

In this \emph{Letter}, we report the results from a search for the stochastic background of GWs induced by scalar-tensor mode couplings of primordial curvature perturbations, in the data from the two LIGO detectors from their O2 run. We assume a log-normal ansatz for the shape of the curvature power spectrum. 
The model parameters are varied 
to span both narrow and broad spectra for which the GW background falls in the LIGO sensitivity band. We find no conclusive evidence of the signals we searched for in the data.


We were therefore able to place upper limits on the amplitude $A$ of the curvature power spectrum using Bayesian parameter estimation. At $95\%$ confidence, the upper limits span $\sim 0.01-0.1$ for a significant fraction of the wavenumber-scales considered. 
This limits the fraction $\beta'$ of the Universe's mass in ultralight PBHs ($M_\mathrm{PBH} \simeq 10^{-20} - 10^{-19} M_{\odot}$) at their formation time to be less than $\sim 10^{-25}$, assuming narrow power spectra ($\sigma \lesssim 0.5$). This means that, even if these black holes exist in the current cosmological epoch (i.e., neglecting Hawking evaporation), they would constitute only a very small fraction of the dark matter ($f_\mathrm{PBH} \lesssim 10^{-15}-10^{-5}$). 

To the best of our knowledge, ours is the first search for the stochastic GW background induced by the primordial curvature perturbations in LIGO data, for mass-scales within the ultralight regime. In addition, it presents the first (GW data-driven) constraints on the amplitude of the curvature power spectrum, and the corresponding upper limits on the PBH abundance, for wavenumber scales spanning $\sim 10^{16} - 10^{18} \mathrm{Mpc}^{-1}$. 
Our upper limits on the PBH abundance are stronger than the existing ones
(derived from the non-observations of the effects caused by the Hawking radiation from PBHs) 
only for a narrow parameter region.
Nevertheless, our GW-based constraints demonstrate that we have finally entered a new era where GW astronomy brings us meaningful information about the extremely small-scale primordial perturbations. In this sense, our results represent a milestone in bridging early-universe cosmology and GW astronomy.

It is almost certain that non-detection of the stochastic GWs originating from the scalar perturbations
by future detectors will tighten the upper limits on the primordial power spectrum and abundance of PBHs
by many orders of magnitude \citep{Inomata2019, Kapadia2020}, thus becoming the most powerful probe of 
the small-scale perturbations.
Since PBH abundance depends quite sensitively on the amplitude of the primordial power spectrum, 
non-detection of such stochastic GWs will completely exclude PBHs in the corresponding PBH mass range 
irrespective of whether they undergo Hawking evaporation or not.
A caveat is that the GW-based constraints on PBHs are indirect and the exclusion of PBHs may be
circumvented if the primordial curvature perturbations are strongly non-Gaussian \citep{Nakama:2016kfq}.

\paragraph{Acknowledgements:} \label{para:ack}

We thank Vuk Mandic for giving valuable feedback on our manuscript. We would also like to acknowledge the Summer School on Gravitational Wave Astronomy (ICTS/gws2019/07) organized by the International Centre for Theoretical Science (ICTS), TIFR, which served as the genesis for this project. SJK’s, KLP’s and PA’s research was supported by the Department of Atomic Energy, Government of India. In addition, SJK’s research was supported by
the Simons Foundation through a Targeted Grant to ICTS. PA’s research was supported by the Max Planck Society through a Max Planck Partner Group at ICTS and by the Canadian Institute for Advanced Research through the CIFAR Azrieli Global Scholars program. TS was supported by the MEXT Grant-in-Aid for Scientific Research on Innovative Areas No. 17H06359, No. 18H04338, and No. 19K03864. Computations were performed using the Alice cluster at ICTS. 
This research has made use of data, software and/or web tools obtained from the Gravitational Wave Open Science Center (https://www.gwopenscience.org), a service of LIGO Laboratory, the LIGO Scientific Collaboration and the Virgo Collaboration. LIGO is funded by the U.S. National Science Foundation. Virgo is funded by the French Centre National de Recherche Scientifique (CNRS), the Italian Istituto Nazionale della Fisica Nucleare (INFN) and the Dutch Nikhef, with contributions by Polish and Hungarian institutes.


\begin{thebibliography}{}
\expandafter\ifx\csname natexlab\endcsname\relax\def\natexlab#1{#1}\fi
\providecommand{\url}[1]{\href{#1}{#1}}
\providecommand{\dodoi}[1]{doi:~\href{http://doi.org/#1}{\nolinkurl{#1}}}
\providecommand{\doeprint}[1]{\href{http://ascl.net/#1}{\nolinkurl{http://ascl.net/#1}}}
\providecommand{\doarXiv}[1]{\href{https://arxiv.org/abs/#1}{\nolinkurl{https://arxiv.org/abs/#1}}}

\bibitem[{Abbott {et~al.}(2019{\natexlab{a}})}]{O2_cross_correlators}
Abbott, B.~P., {et~al.} 2019{\natexlab{a}}.
\newblock \url{https://dcc.ligo.org/LIGO-T1900058/public}

\bibitem[{Abbott {et~al.}(2019{\natexlab{b}})}]{O2_stochastic_search}
---. 2019{\natexlab{b}}, Phys. Rev. D, 100, 061101,
  \dodoi{10.1103/PhysRevD.100.061101}

\bibitem[{Aghanim {et~al.}(2018)}]{Aghanim:2018eyx}
Aghanim, N., {et~al.} 2018.
\newblock \doarXiv{1807.06209}

\bibitem[{Allahverdi {et~al.}(2020)}]{Allahverdi:2020bys}
Allahverdi, R., {et~al.} 2020.
\newblock \doarXiv{2006.16182}

\bibitem[{Allen \& Romano(1999)}]{Allen1999}
Allen, B., \& Romano, J.~D. 1999, Phys. Rev. D, 59, 102001,
  \dodoi{10.1103/PhysRevD.59.102001}

\bibitem[{Ananda {et~al.}(2007)Ananda, Clarkson, \& Wands}]{Ananda:2006af}
Ananda, K.~N., Clarkson, C., \& Wands, D. 2007, Phys. Rev. D, 75, 123518,
  \dodoi{10.1103/PhysRevD.75.123518}

\bibitem[{{Bugaev} \& {Klimai}(2011)}]{Bugaev2011}
{Bugaev}, E., \& {Klimai}, P. 2011, \prd, 83, 083521,
  \dodoi{10.1103/PhysRevD.83.083521}

\bibitem[{{Bugaev} \& {Klimai}(2010)}]{Bugaev2010}
{Bugaev}, E.~V., \& {Klimai}, P.~A. 2010, Soviet Journal of Experimental and
  Theoretical Physics Letters, 91, 1, \dodoi{10.1134/S0021364010010017}

\bibitem[{{Carr} {et~al.}(2020){Carr}, {Kohri}, {Sendouda}, \&
  {Yokoyama}}]{Carr2020}
{Carr}, B., {Kohri}, K., {Sendouda}, Y., \& {Yokoyama}, J. 2020, arXiv
  e-prints, arXiv:2002.12778.
\newblock \doarXiv{2002.12778}

\bibitem[{Carr {et~al.}(2017)Carr, Raidal, Tenkanen, Vaskonen, \&
  Veerm\"ae}]{Carr1_2017}
Carr, B., Raidal, M., Tenkanen, T., Vaskonen, V., \& Veerm\"ae, H. 2017, Phys.
  Rev. D, 96, 023514, \dodoi{10.1103/PhysRevD.96.023514}

\bibitem[{Carr {et~al.}(2010)Carr, Kohri, Sendouda, \& Yokoyama}]{Carr2_2017}
Carr, B.~J., Kohri, K., Sendouda, Y., \& Yokoyama, J. 2010, Phys. Rev. D, 81,
  104019, \dodoi{10.1103/PhysRevD.81.104019}

\bibitem[{{Chen} {et~al.}(2020){Chen}, {Yuan}, \& {Huang}}]{Nanograv}
{Chen}, Z.-C., {Yuan}, C., \& {Huang}, Q.-G. 2020, \prl, 124, 251101,
  \dodoi{10.1103/PhysRevLett.124.251101}

\bibitem[{{Chluba} {et~al.}(2012){Chluba}, {Erickcek}, \&
  {Ben-Dayan}}]{Chluba2012}
{Chluba}, J., {Erickcek}, A.~L., \& {Ben-Dayan}, I. 2012, \apj, 758, 76,
  \dodoi{10.1088/0004-637X/758/2/76}

\bibitem[{Chluba {et~al.}(2019)}]{Chluba:2019kpb}
Chluba, J., {et~al.} 2019, Bull. Am. Astron. Soc., 51, 184.
\newblock \doarXiv{1903.04218}

\bibitem[{Christensen(1992)}]{Christensen1992}
Christensen, N. 1992, Phys. Rev. D, 46, 5250, \dodoi{10.1103/PhysRevD.46.5250}

\bibitem[{Christensen(2018)}]{Christensen2018}
---. 2018, Reports on Progress in Physics, 82, 016903,
  \dodoi{10.1088/1361-6633/aae6b5}

\bibitem[{{Fixsen} {et~al.}(1996){Fixsen}, {Cheng}, {Gales}, {Mather},
  {Shafer}, \& {Wright}}]{Fixsen1996}
{Fixsen}, D.~J., {Cheng}, E.~S., {Gales}, J.~M., {et~al.} 1996, \apj, 473, 576,
  \dodoi{10.1086/178173}

\bibitem[{{Hawking}(1975)}]{Hawking1975}
{Hawking}, S.~W. 1975, in Quantum Gravity, 219--267

\bibitem[{Inomata {et~al.}(2016)Inomata, Kawasaki, \& Tada}]{Inomata:2016uip}
Inomata, K., Kawasaki, M., \& Tada, Y. 2016, Phys. Rev. D, 94, 043527,
  \dodoi{10.1103/PhysRevD.94.043527}

\bibitem[{Inomata \& Nakama(2019)}]{Inomata2019}
Inomata, K., \& Nakama, T. 2019, Phys. Rev. D, 99, 043511,
  \dodoi{10.1103/PhysRevD.99.043511}

\bibitem[{{Inomata} \& {Terada}(2020)}]{InomataTerada2020}
{Inomata}, K., \& {Terada}, T. 2020, \prd, 101, 023523,
  \dodoi{10.1103/PhysRevD.101.023523}

\bibitem[{Jeong {et~al.}(2014)Jeong, Pradler, Chluba, \&
  Kamionkowski}]{Jeong:2014gna}
Jeong, D., Pradler, J., Chluba, J., \& Kamionkowski, M. 2014, Phys. Rev. Lett.,
  113, 061301, \dodoi{10.1103/PhysRevLett.113.061301}

\bibitem[{Kapadia {et~al.}(2020)Kapadia, Pandey, Suyama, \&
  Ajith}]{Kapadia2020}
Kapadia, S.~J., Pandey, K.~L., Suyama, T., \& Ajith, P. 2020, Phys. Rev. D,
  101, 123535, \dodoi{10.1103/PhysRevD.101.123535}

\bibitem[{Kohri \& Terada(2018)}]{Kohri2018}
Kohri, K., \& Terada, T. 2018, Phys. Rev. D, 97, 123532,
  \dodoi{10.1103/PhysRevD.97.123532}

\bibitem[{Lyth \& Liddle(2009)}]{Liddle-Lyth}
Lyth, D., \& Liddle, A. 2009, The primordial density perturbation (Cambridge
  University Press)

\bibitem[{Mandic {et~al.}(2012)Mandic, Thrane, Giampanis, \&
  Regimbau}]{Mandic2012}
Mandic, V., Thrane, E., Giampanis, S., \& Regimbau, T. 2012, Phys. Rev. Lett.,
  109, 171102, \dodoi{10.1103/PhysRevLett.109.171102}

\bibitem[{Matarrese {et~al.}(1994)Matarrese, Pantano, \&
  Saez}]{Matarrese:1993zf}
Matarrese, S., Pantano, O., \& Saez, D. 1994, Phys. Rev. Lett., 72, 320,
  \dodoi{10.1103/PhysRevLett.72.320}

\bibitem[{Nakama {et~al.}(2014)Nakama, Suyama, \& Yokoyama}]{Nakama:2014vla}
Nakama, T., Suyama, T., \& Yokoyama, J. 2014, Phys. Rev. Lett., 113, 061302,
  \dodoi{10.1103/PhysRevLett.113.061302}

\bibitem[{Nakama {et~al.}(2016)Nakama, Suyama, \& Yokoyama}]{Nakama:2016kfq}
---. 2016, Phys. Rev. D, 94, 103522, \dodoi{10.1103/PhysRevD.94.103522}

\bibitem[{{Romano} \& {Cornish}(2017)}]{Romano2017}
{Romano}, J.~D., \& {Cornish}, N.~J. 2017, Living Reviews in Relativity, 20, 2,
  \dodoi{10.1007/s41114-017-0004-1}

\bibitem[{Saito \& Yokoyama(2009)}]{Saito:2008jc}
Saito, R., \& Yokoyama, J. 2009, Phys. Rev. Lett., 102, 161101,
  \dodoi{10.1103/PhysRevLett.102.161101}

\bibitem[{{The LIGO Scientific Collaboration} {et~al.}(2019){The LIGO
  Scientific Collaboration}, {the Virgo Collaboration}, {Abbott},
  {et~al.}}]{GW_Open_Data}
{The LIGO Scientific Collaboration}, {the Virgo Collaboration}, {Abbott}, R.,
  {et~al.} 2019, arXiv e-prints, arXiv:1912.11716.
\newblock \doarXiv{1912.11716}

\bibitem[{Tomita(1967)}]{Tomita:1967sv}
Tomita, K. 1967, Prog. Theor. Phys., 37, 831, \dodoi{10.1143/PTP.37.831}

\bibitem[{Wang {et~al.}(2019)Wang, Terada, \& Kohri}]{Wang2019}
Wang, S., Terada, T., \& Kohri, K. 2019, Phys. Rev. D, 99, 103531,
  \dodoi{10.1103/PhysRevD.99.103531}

\bibitem[{Wang {et~al.}(2018)Wang, Wang, Huang, \& Li}]{Wang2018}
Wang, S., Wang, Y.-F., Huang, Q.-G., \& Li, T. G.~F. 2018, Phys. Rev. Lett.,
  120, 191102, \dodoi{10.1103/PhysRevLett.120.191102}

\end{thebibliography}


\end{document}